# Interlaminar toughening in structural carbon fiber/epoxy composites interleaved with carbon nanotube veils

Yunfu Ou[1,2], Carlos González[1,2], Juan José Vilatela[1]

[1] IMDEA Materials Institute

C/ Eric Kandel 2, 28906 Getafe, Madrid, Spain

[2] E. T. S. de Ingenieros de Caminos, Universidad Politécnica de Madrid, 28040 Madrid, Spain.

**Abstract**

The susceptibility to delamination is one of the main concerns in fiber reinforced polymer composites (FRPs). This work demonstrates improvements of 60% in Mode-I fracture toughness after integration of thin (~30 micron), continuous veils of carbon nanotubes (CNTs) directly deposited onto carbon fiber fabric as the CNT are drawn from the gas-phase using a semi-industrial process. A combination of optical imaging, scanning electron microscopy and a Raman spectroscopy provide a new rapid tool to unambiguously determine the crack propagation path by simple visual inspection of fracture surface. The results show that interlaminar crossing between CNT veil/CF interfaces is of paramount importance. The crack front alternatingly propagates above and below the CNT-toughened interlayer, significantly improving the fracture toughness of resultant laminates. This mechanism is strongly influenced by the method used to integrate the veils onto the CF. CNT veils directly deposited onto the fabrics as a low-density layer lead to large improvements in interlaminar properties, whereas compact CNT veils densified by solvent exposure prior to their integration in the lay-up act as defects.

# 1. Introduction

The past half-century has witnessed the rapid expansion of fiber reinforced polymer (FRP) composites. Their higher specific stiffness, strength and outstanding fatigue properties combined with environmental and chemical resistance make them ideal candidates for thousands of structural applications in aircraft, automobiles, civil applications, ships and offshore platform facilities, amongst others [1-4]. These high performance structural composites are based on lay-ups of stacked plies (e.g. prepregs or dry wovens) leading to extraordinary in-plane properties, dominated by fiber properties, but suffering from much poorer out-of-plane properties, which are strongly matrix-dependent. This last effect is particularly problematic and failures due to matrix cracking and ply delaminations often occur by the application of out-of-plane loads exerted during impact events [5]. There have been extensive efforts to improve interlaminar properties of structural composites, including strategies such as matrix toughening (rubber[6], thermoplastic[7] etc.), hybridization[8], stitching [9], 3D weaving [10] and Z-pinning [11] etc. These methods, however, may lead to increase in cost, weight, or loss of in-plane properties. An approach that is attractive to alleviate the above problems is by fiber interleaving [4, 12-17].

Interleaving is a common strategy used in the aircraft industry to enhance acoustic damping and interrupt fatigue crack propagation in metallic structures [18]. It should be mentioned that the interleave concept using micro-length or short fiber has been tested as early as 1994 [12]. Furthermore, the concept has been extended from the mid-layer (research only) to multiple layered reinforcement (practical pre-preg applications) [17, 19].

Very recently, electrospun thermoplastic nanofiber interleaves were used to improve the interlaminar fracture toughness (ILFT) of composite [4, 13-15, 20-24]. Their nanoscale diameter offers the possibility of very thin interlayers, which offers the opportunity to reinforce the interlaminar bonding with minimal weight penalty and effectively avoids reductions in in-plane properties. Furthermore, because of their high porosity, the resin flow is hardly affected, which is influential prerequisite of better infiltration. Several experiments focused on utilizing electrospun nanofibers which can be dissolved in the resin such as Polyetherketone cardo (PEK-C) [22], polysulfone (PSF) [25] or phenoxy-based nanofibers [26]. The reported improvements in IFT would be ascribed to a phase separation of the dissolved nanofibers which leads to tough particulate phases[25]. Several polymer types which can retain their nanofibrous structure in the final composite, such as polyacrylonitrile (PAN) [27], polyimide (PI) [28], and Nylon 6.6 [29, 30] etc., have already been investigated. The advantages of keeping the nanofibrous structure are that the physical and mechanical properties of nanofibers can be retained and that there is no increase in viscosity of the matrix resin due to dissolution of the nanofibers.

Compared with electrospun nanofibers, CNTs seem to be more promising. Firstly, the diameter of CNTs is generally smaller than electrospun nanofibers, which enables thinner interleaves. Second, CNT possess higher mechanical properties (tensile strength, Young's modulus et al.) and larger specific surface area, thus enable to build a stronger mechanical link between the fracture surfaces. Last but not least, exceptional electrical (when metallic[31]) and thermal conductivity of CNT endow resultant composites multifunctionalities, such as prevention of unwanted electrostatic discharges, electromagnetic interference shielding or damage suppression during lightning strike and power management [32] as well as potential for damage sensing of hierarchical composites [33].

Several approaches for integrating CNTs into FRP structural composites have been studied [34-37]. Directly mixing CNTs into low viscosity epoxy resin has been a popular route to produce hierarchical CNT-FRP hybrid, but the dramatic increase in viscosity upon addition of CNT and filtering effects by fiber preforms during liquid moulding processes often lead to poor CNT dispersion and incomplete textile impregnation [38]. A compression moulding route using short CNTs and reducing resin flow distance has partially solved this challenge and lead to increases in fracture energy of 24% [39]. Using CNT as sizing material was reported by Warrier et al.[36], it showed that the presence of CNTs in the sizing resulted in an increased resistance of crack initiation fracture toughness by +10%, but a lowered crack propagation toughness of −53%. Another popular route has consisted in the growth of CNTs on the surface of CF [40]. After improvements in the synthesis method to avoid degradation of the CF by the metallic catalyst, interlaminar properties have been improved while preserving longitudinal composite properties [41]. A spray coating technique [42] has been used as an effective way to deposit CNTs onto prepregs with good control of network formation and the potential for localization. The resultant laminates demonstrated both improved fracture toughness and integrated damage sensing capability. These studies have helped identify the potential of CNT-based interleaves for interlaminate reinforcement, but the use of cumbersome laboratory processes to integrate the CNT powder in the composite structure, limits their widespread implementation.

More recently, there has been emerging interest in the use of thin, pre-fabricated porous unidirectional fabrics of CNT fibers (veils) as interleaves in hierarchical CNT/CF/polymer composites. Using pre-formed fabrics directly spun from a CNT aerogel produce by floating catalyst chemical vapour deposition (FC-CVD [43]), avoids the well-known challenges related to increased viscosity when using nanocarbons as fillers, or degradation in fibers properties when CNTs are directly grown on the CF surface [44]. The integration of pre-assembled CNTs without recourse to solvents is also potentially compatible with standard composite fabrication methods and essentially analogous to the introduction of thermoplastic veils already used industrially [14]. Furthermore, CNT "veil" fabrication processes on semi-industrial scale have successfully demonstrated, with even laboratory equipment producing tens of kilometers per day of this material [45], and with CNT fibers from the same veils, outperforming reference high-performance fibers [46]. Also very importantly, CNT veils have reproducible, controlled mechanical properties, enabling identification of clear structure-property relations of general applicability to other nanomaterials and across different laboratories and industrial facilities. As a final motivation for the study of CNT veils, we point out that the FC-CVD production of CNT veils is inherently similar to carbon black (< 0.1 €/kg) synthesis, whereby an inexpensive carbon source (e.g. methane, natural gas) is converted into CNT veils in a single step, with a yield above 15% and spinning rates as high as 100m/min. Thus, chemical engineering analysis indicates that both the cost and greenhouse gas emissions of CNT fiber production will be well below that carbon fiber, for example.

In spite of these credentials, there has been little work in producing hierarchical composites with CNT fiber veil interleaves for interlaminar reinforcement. Song et al. [47] interleaved a chemically post-processed CNT sheet into fiberglass plies utilizing layer by layer method. Results showed that SBS strength increased by 9.1%. Nguyen et al. [48] introduced aligned CNT sheet into unidirectional CFRP system. Mode II interlaminar fracture toughness was found to be significantly improved when interlayer tougheners were combinedly used, no increments in Mode I fracture toughness were reported.

Overall, the evidence of these two reports is that the same veil material can lead to different interlaminar properties depending on factors including morphology, method of integration in the laminate and composite consolidation conditions. Thus, it is critical not only to produce improvements in interlaminar properties, but also to rationalize them in terms of these factors. In this paper, we introduce a new strategy to integrate CNT veils to woven carbon fiber fabrics/epoxy system to produce structural laminate composites with improved interlaminar properties. The veils are directly affixed to the surface of carbon fabric at the point of fabrication and do not require any previous treatment. Composite specimens for mode-I delamination tests are produced under optimized Vacuum Assisted Resin Transfer Molding (VARTM). The results show that the degree of compaction of the starting CNT fiber veil is the dominant factor with respect to the critical strain energy release rate in mode-I delamination ($G_{IC}$). Taking advantage of the strong resonant Raman scattering signal from the CNTs synthesized in this work, we could determine the crack propagation path and used it as a fractography method to describe the observed $G_{IC}$ values.

## 2. Experimental

2.1 Materials

Macroscopic veils of CNT fibers were synthesized by the direct spinning process, whereby a CNT web (aerogel) is directly drawn from the gas-phase during growth of CNTs by FC-CVD. The CNTs were synthesized under using ferrocene as iron catalyst, thiophene and butanol as sulfur catalyst promoter and carbon source, respectively, in the hydrogen atmosphere at 1250 °C with precursor feed rate of 5 mL h$^{-1}$ and a winding rate of 5 m min$^{-1}$. These conditions produce long, highly graphitic CNTs with 3-5 layers [45] and a low degree of alignment in the aerogel [49]. The diameters of CNTs range from 3-9 nm [45]. The FC-CVD reaction can be conducted using methane, or even natural gas, as carbon source, which could make it more sustainable than polymer interleaves or even the CF itself. As shown in **Fig. 1a** the CNT fibers were continuously drawn out at the exit of the furnace and directly deposited onto the surface of a piece of carbon fabric which was wrapped on a winder of D = 10 mm diameter. Winding time of CNT fibers was set as 0.5 h, which gave an areal density of 0.8 g/m$^2$, which results in a thickness of 30 micron of CNT veil after infusion (in its as-made state the same veil was fluffy and would have a thickness of more than 3 mm). This thickness was chosen for two reasons: a) the interleaf is thin relative to a CF ply, but sufficiently thick to study possible cohesive failure through the interleaf, and b) the thickness is similar to that found in commercial veils produced by companies that have scaled up this FC-CVD process.

The carbon fabrics used in this study was the G0926 Hexcel woven fabric with a weave style of 5H satin and 370g/m$^2$ of areal weight (0.38mm of ply thickness). The final laminate contained eight carbon plies [0°$_4$-CNT$_{veil}$-0°$_4$] with the CNT interleave in the mid-plane. The preform was produced by means of two 4-layer halves with the CNT film deposited on the subsequent mid-plane of the laminate. DERAKANE 8084 was chosen as matrix which is an elastomer-modified bisphenol-A epoxy vinyl ester resin for structural applications and provided by Ashland Inc. Both the MEKP hardener and Cobalt octoate catalyst were available from Plastiform S.A. By using a recommended concentration of precursors (100:1.5:0.3), the mixture will reach its gel point after 30-60 min.

2.2 Laminate preparation

The interleaves were produced by winding continuous fiber of CNTs directly onto CF fabric wrapped around a cylindrical winder (**Fig. 1**). Two layers of infusion media were placed onto the upper and bottom surfaces to promote the resin flow, to guarantee complete wet-out of the preform and to eliminate dry spots and voids. This method ensures good impregnation in those regions affected by the teflon film (25 μm) insertion used for crack initiation in fracture coupons. The length of infusion media was designed to be 30 mm shorter than that of fabric in order to slow down the speed of flow front before it reaches the outlet and leave enough time to complete infiltration of z-direction. Peel plies were purposely inserted into the interfaces between infusion media and carbon fabrics for the easiness of removing the redundant parts after curing. Before starting infusion, resin was degassed for 10 minutes in a vacuum chamber under a vacuum degree of -0.88 bar. After completing the infusion process, the inlet and outlet of the system was closed off and cured at room temperature (RT) for 24 hours, followed by a post curing step at 60°C for 2 hours. Note that CNT-interleaved composites and baseline (without CNT) were manufactured in a same batch to minimize the influence of other uncontrollable factors on the testing results.

2.3 Double cantilever beam (DCB) tests

Among the three independent fracture modes, i.e. the crack opening mode (mode I), the in-plane shear mode (mode II) and out-plane shear mode (mode III), Mode I is regarded as more important than other two because the its fracture toughness is usually smaller than those of mode II and mode III and so the fracture is easily initiated and propagated under the mode I loading condition [50]. Out of this reason, we mainly focused on Mode I fracture toughness in this paper.

*2.3.1 DCB specimen preparation*

The standard DCB specimens were cut from aforementioned well-prepared laminates utilizing a water-refrigerated milling machine. The lateral side of each specimen was then sprayed with white primer paint and the 60 mm after the pre-crack marked in 1 mm increments to track crack growth throughout the test. The configuration and size of the standard specimen can be seen in **Fig. 2**.

*2.3.2 Test procedure*

The DCB tests were carried out in accordance with ASTM Standard D5528-01 [51], which specifies either hinges or end blocks for load introduction. Measurements were conducted in a screw-driven testing machine (Instron 3384) at cross-head spead of 1 mm/min. Force was recorded by utilizing a 500 N load cell and the DCB arm opening displacement using the cross-head movement. Crack growth was visually observed every millimeter and listed in an experimental sheet. Time was also recorded in order to correlate the crack growth with the force and displacement. In order to assure the crack propagation, specimens were pre-loaded to create a ~5 mm pre-crack. After unloading, the specimen was loaded again until another 50 mm propagation had been reached.

*2.3.3 Data reduction*

There are several data reduction methods can be employed to calculate the Mode I fracture toughness ($G_{IC}$) involving Area method, Beam Theory (BT) method, Modified Beam Theory (MBT) method, Compliance Calibration (CC) method and Modified Compliance Calibration (MCC) method[51, 52]. In our work, MBT method was chosen because it yields the most conservative results [51]. Traditional BT method without correction overestimates the Mode I fracture toughness value because of the imperfect beam built-in during the DCB test, which allows the rotation at the crack tip. This overestimation can be balanced by pretending the DCB specimen has a slightly longer delamination so the crack total length is replaced by $(a + |\Delta|)$, the strain energy release rate ($G_I$) can be obtained by the following equation:

$$G_I = \frac{3 * F * \delta}{2 * w * (a + |\Delta|)}$$

where F is the load; δ is the load point displacement; w is the sample width, and a is the crack length from the load point. The coefficient $\Delta$ could be determined by plotting the cube root of the compliance (displacement divided by force) against the crack length.

Data points from DCB experiments were used to construct a delamination resistance curve (*R*-curve), which is the graph of strain energy release rate (*G*) versus crack length (a). From the *R*-curve, two types of critical *G* values can be calculated, the initiation ($G_{IC,ini}$) and propagation ($G_{IC,prop}$) values. $G_{IC,ini}$ is the energy dissipated when the crack starts to propagate, which might vary for each specimen as the value strongly depends on the crack tip condition [53] and location of the edge of the film [54]. $G_{IC,prop}$ is the value where the *G* value achieve a constant state and the crack propagates in a self-similar way independent on the crack length. The *G* value of interest for this experiment was $G_{IC,prop}$, because it is more realistic in practice for determining the likelihood of the crack to continue to propagate.

2.4 Other Characterization techniques

C-Scan was performed to determine the quality of the panel and to localize manufacturing defects. Scanning electron microscopy (EVO MA15, Zeiss) and FIB-FEGSEM microscope (Helios NanoLab 600i FEI) were used to investigate the morphology of CNT fiber veils and the fracture surfaces of the laminates. An optical microscope (OLYMPUS BX51) was used to visualize the distribution of CNT veils in the cross section of the composite laminates and then to determine the thicknesses of the CNT layer. Raman spectroscopy (Renishaw PLC) was performed with an excitation wavelength of 532 nm from a tunable YAG laser focused on the sample using laser power in the range of 1-5 mW so as to avoid sample heating.

## 3. Results and discussion

3.1 Optimized composite fabrication and structural characterization

An optimized VARTM method was employed to make the CF/CNT veils/epoxy hierarchical laminate (**Fig. 1**). Specifically, CNT fiber veils were deposited onto carbon fabric by winding them directly around the fabric at the point of spinning from an aerogel. The resulting veil material is a highly porous, thin, nonwoven fabric of interconnect CNTs which is consolidated in the composite as it is pressed under vacuum pressure used during composite compaction. The veil forms a thin film strongly attached to the CF fabric surface. Even though the VARTM method is a very mature method in the area of laminate manufacturing, the production of high-quality and almost void-free composites requires good control of multiple factors. For the fabrication of these samples, we found a long infusion time particularly beneficial to ensure infiltration of resin through the composite thickness and the veil. The optimized VARTM protocol included using 2 layers of distribution medium (**Fig. 1c**), 10 minutes of resin-degassing and 20 minutes of infusion time. These conditions lead to uniform and complete resin flow across the laminate (**Fig. 1d**).

The composite panels were 30 × 30 cm, and purposely designed to comprise an interleaved area of 12 × 30 cm and the remaining part free of interleaf. Reference interleaf-free DCB specimens produced under identical conditions could thus be cut off from the same panel, although in general the optimized fabrication process produced small variation between batches. **Fig. 3a** shows an example of a final CF/CNT veils/epoxy hybrid composite produced using the aforementioned optimized method. Visual inspection proved that there exist no obvious defects/voids on its surface. C-Scan picture (**Fig. 3b**) combined with cross sectional micrographs (**Fig. 3c** and **d**) further confirmed almost void-free laminate production. Optical microscopy (**Fig. 3d**) also shows that the CNT veil maintains its shape after infusion and is fully integrated in the composite. It follows the contour of the CF fabric leaving no gap between the two plies, which is a prerequisite to reinforce the interface between the two laminae. SEM characterization of the interleaf region (after laminate fracture), as shown in the example in **Fig. 3e,** confirmed that the veils are successfully infiltrated by the matrix.

The interleaf thickness determined from optical micrographs is around 30 microns and corresponds to a volume fraction of approximately 1%. It was controlled by adjusting the time the CNT fiber was wound on the spool. At its exit from the CVD reactor and deposition on the CF fabric the CNT veil has a very low density (< 0.02 g/cc) reminiscent of the CNT aerogel morphology. For reference, in its as-made state the same veil would have a thickness of more than 3 mm. Densification of CNT fibers by capillary forces induced by exposure to volatile solvents such as ethanol and acetone is known to increase longitudinal tensile properties [55, 56] and can improve adhesion to surfaces. Understanding the role of such densification is not only important to establish clear veil structure-interlaminar property relations, but also because commercial larger scale CNT veils are available already in densified format (they could be handled otherwise). In order to gain a first insight into the role of densification we produced composites of two types. One had CNT veils as-produced, and in the other the CNT veils were densified at the point of spinning by exposure to ethanol spray.

3.2 Mode I interlaminar fracture toughness

Pristine and CNT veil-interleaved woven composites were produced with the 5H Satin weave fabrics as described and tested as discussed above. Though the load-extension curves of woven composites always display stick-slip crack propagation behavior, as shown in **Fig. 4a**, the data can still be utilized to generate R-curves [57]. There is a small scatter of the $G_{IC}$ values in the resulting R-curves (see **Fig. 4b**) related to the local variation of toughness on account of the interlaminar resin pockets and the wavy nature of the woven fabric, but each groups has clearly distinctive values.

There is moderate fiber bridging in the pure woven laminates, resulting in a relatively flat R-curve, which is in line with similar systems [13]. For the control samples without CNT veils, the difference between $G_{IC,ini}$ and $G_{IC,prop}$ is marginal, especially when compared with the differences observed in unidirectional fiber reinforced composites [13, 58]. Incorporation of the low-density, as-produced CNT veils, produces a sharp rise in the R-curve, with the propagation value of fracture toughness ($G_{IC,\,prop}$) increasing up to ~50% compared with the initiation value ($G_{IC,\,ini}$) due to a hierarchical fiber-bridging toughening mechanism (see **Section 3.3**). Remarkably, the average values of $G_{IC,\,prop}$ for as-produced CNT samples are 60% higher compared with the corresponding value of the baseline material. In contrast, the densified veils produced a 28.4% decrease of $G_{IC,\,prop}$. This layer functioned essentially as a defect that weakened the interlaminar region and did not produce any fiber-bridging, thus effectively reducing the fracture toughness of resulting laminate.

3.3 Toughening mechanisms

*3.3.1 Fractography*

The fracture surfaces of tested specimens were examined by electron microscopy and Raman spectroscopy. SEM micrographs of the reference specimen (**Fig. 5a**) show a clean matrix fracture surface with no fiber-bridging, in line with its flat R-curve shape. These are typical features of low toughness mode-I fracture surfaces. The fracture surface is totally different (see **Fig. 5b**) for the composite with as-produced CNT veils. Overall, the fracture surface is now quite rough, which is beneficial for energy dissipation and therefore as a toughening mechanism. Closer inspection shows that there are two levels of fiber bridging operating at different length scales: firstly, the presence of broken carbon fibers evidences microscale carbon fiber bridging and pull-out, and the presence of substantial of pull-out CNTs indicates matrix level CNT bridging. These combined mechanisms lead to a rising R-curve.

In comparison, the fracture surface of laminates interleaved with densified CNT veils (**Fig. 5c**) shows peeling of the CNT interleaf layer. No carbon fiber was easily detected at fracture surface, which indicates that the densified CNT layer guided the crack and hindered the formation of carbon fiber bridging. This is in agreement with the flat R-curve obtained for these specimens, and hence, the fracture energy obtained in this case can be endorsed to the intrinsic fracture energy of the CNT layer in the absence of other superior toughening mechanisms.

*3.3.2 Crack propagation*

Toughening effect, to a considerable degree, depends on crack behavior. Several surface analysis technologies were used in order to get a deeper insight into the crack initiation and propagation behaviors as well as the corresponding toughening mechanisms. The starting point of this analysis is the observation that the fracture surfaces of the two types of samples with CNT interleaves exhibit clear differences by simple visual inspection. The tougher one, with the low-density interleaf, consists of a complex pattern of "patches" of three different, well-defined colours (shiny - S, grey - G and dark black - B). Examples of these regions are marked in the fracture surfaces of **Fig. 6**. A further important observation is the composition of the complementary fracture surfaces of the two arms in these samples. The position of black regions in one arm (B) always corresponds to black regions in the other arm (B), and vice versa, while Grey regions (G) are matched by shiny regions (S).

The composition of the different regions could be readily determined from their Raman spectra, taking advantage of the strong Raman intensity and distinctive features of the CNT veils compared with the CF. Considering the low D/G ratio and sharp G band lineshape, grey and dark-black regions can be easily recognized as CNT-rich regions, whereas the shiny areas, with broad D and G bands of similar intensity, correspond to carbon fiber. Thus, the Raman data enable to identify different colors on the fracture surface to different layers of the composite and ultimately, to different interfaces: S/G represent CF/CNT veil interfaces and B/B represent CNT veil/CNT veil interfaces.

**Fig. 7** shows SEM micrographs of the different region and the crack propagation path derived from the fractography analysis. As shown in the picture, the shiny region (**S**) consists of pure carbon fibers. Those fibers debonded from CNT-rich resin areas and left substantial dents on the counterpart (**G**). Thus it is of reasonable to conclude that the crack moved forward through the interface of carbon fabric layer and CNT layer, as schematically shown in **Fig. 7 (II)**. The dark-black (**B**) region in one arm always correspond to a dark-black region (**B**) in its counterpart (see **Fig. 6**), and SEM micrograph (**Fig. 7 B**) indicates that both regions shared the similar fracture morphologies with extensive CNTs. The crack, in this case, moves exclusively though the CNT layer, as schematically depicted in **Fig. 7 (I).** Following this logic, the propagation route of crack in whole area of interest could be unveiled simply based on the apparent color of fracture surfaces.

Equipped with this fractography method the differences in fracture toughness between the two types of CNT-containing samples can be rationalized in terms of their microstructure, and traced back to their fabrication method. **Fig. 8** presents the fracture surfaces of two different samples and schematics of the different crack propagation paths. In the low toughness samples the crack progresses almost exclusively along the CNT-rich interlayer, corresponding to a cohesive failure. A similar behavior is observed in composites produced with CNT veils produced semi-industrially, which are supplied as densified free-standing sheets [59]. In contrast, the high toughness CNT interleaf sample showed regular crossings of the interlaminar region through the interlayer. Essentially a combination of adhesive and cohesive failure. These saw-like interlaminar crossings not only increase the total crack propagation length, but also trigger different toughening mechanisms: crack advance through the nanotoughened epoxy, possibly nano-scale bridging, and a good deal of carbon fiber bridging (micro scale). Thus, failure in this mode combines pull-out of nano-sized reinforcements, debonding between carbon fiber and resin, as well as carbon fiber pull-out and breakages (see **Fig. 5b** and **Fig. 7S/G**). These mechanisms are usually considered to consume additional energy and hence contribute significantly to the enhanced interlaminar toughness and thus Mode I fracture behavior.

The results presented above show evidence of interlaminar reinforcement by adding CNT veils when they are appropriately synthesized and integrated in the laminate. Densification of the CNT veil appears as a critical processing parameter. It is expected that the structure resulting from the compaction of the dry veil has lower density than the veils under capillary-induced solvent densification. Upon resin infiltration the resulting composite interleaf is expected to have different CNT volume fraction and properties such as intrinsic fracture toughness and in general different cohesion in the transverse direction. There is ample work showing the enhancement of CNT fiber properties upon polymer infiltration, with the matrix improving stress transfer between CNT bundles and leading to a complex hierarchical composite structure [60-63]. Further studies on the fracture toughness of CNT fiber interleaves and their composites should shed more light into the relationship between veil properties and interlaminar reinforcement.

## 4. Comparison on the state of the art

Over the past decades, huge effects have been paid for the improvement of interlaminar fracture toughness in fiber/epoxy composites through nanoparticle reinforcement, as summarized in **Table 1**. These works mainly vary with respect to the filler integration method, fabrication method adopted, composite system, mass fraction of nano-fillers, fiber geometry and configuration. Generally, the major CNT integration techniques can be classified and broken down into: (i) mixing in bulk resin [64-68], (ii) in-situ CNT growth (i.e. CNT sizing)[40, 69, 70], (iii) physical transfer of the CNT particles (i.e. spray coating) [42, 71-75] and (iv) CNT interleaving technique [13, 24, 48, 76]. The first method often leads to large improvements in matrix properties at low mass fractions, however, due to issues such as agglomeration and filtering effects of nanofillers, improvements in mechanical properties under higher mass fractions were rarely reported. In-situ growth of CNTs on the carbon fiber surfaces can substantially improve the fracture toughness of the resulting composite more than 50% [40, 69, 70]. Whilst this technique overcomes the puzzle of unwanted high melt viscosity, the high growth temperature, which is typically associated with CNT growth (600 –1200 °C), leads to mechanical degradation for the CFs and removing of the polymer sizing [77]. Spray coating allows for the creation of truly nano-engineered hierarchical composite systems with tailored CNT localization for improved fracture toughness [42]. The main drawback of this method is the CNT should firstly dispersed in solvent [42, 71-75] and the leftover of those solvent may be harmful to the mechanical properties of resultant laminate. Interleaving nanofiber assemblies (electrospun thermoplastic fiber veils or CNT fiber veils) is particularly competitive compared with above mentioned methods as these veils can be easily and readily deposited in between the primary reinforcing fiber layers before infusion, thus the viscosity of the composite matrix is not affected. Additionally, their nanoscale diameters enable the production of ultra-thin interleaves, which can be utilized to improve the interlaminar fracture toughness with minimal weight penalty. It is worth noting that integration of CNT fiber veil, compared with electrospun thermoplastic fiber veil, would be more attractive since the former also endow the composite structure multifunctionalities (i.e. SHM [33]).

In short, CNT veil based approaches provide completely new fields for the application of CNT with advantages not only simplifying the integration process but also significant mechanical enhancement. Next to the highly effective use of nanofiller reinforcement materials, it bears the feasibility of scale-up for real industrial manufacturing.

## 5. Concluding remarks

This work provides new insights into the micro-mechanisms of interlaminar toughening in structural composites with woven CF fabrics interleaved with CNT fiber veils. A facile and scalable infusion protocol was successfully developed to make CNT veil/CF/epoxy hybrids, leading to void-free laminates manufactured by VARTM without utilizing additional expensive or complex techniques. The effects of CNT veils on the interlaminar fracture toughness are investigated and discussed comparatively, followed by a systematic analysis of toughening mechanisms. These results demonstrate that the degree of densification of the CNT veil exerts a profound influence on the resulting interlaminar properties. Low-density "fluffy" CNT veils consistently led to CFRP laminates with as much as 60% enhancement of $G_{IC,\ prop}$. In contrast, integrating densified CNT veils deteriorated their interlaminar mechanical performances.

A simple new methodology combining Raman analysis and direct optical observation was used to study crack propagation in different samples. Effectively, Raman spectroscopy can be used as a convenient and powerful tool for fractography analysis by unambiguously providing compositional information over large areas of the fracture surface and relating it directly to the color of the distinctive regions. The analysis results showed that interlaminar crossing plays a dominant roles amongst toughening mechanisms. The crack front propagates alternatingly between interfaces of the laminate, triggering multi-level fiber bridging and significantly improving the fracture toughness of the laminate. The behavior observed in these composites cannot be generalized to other composite architectures without caution, but it is expected that the analysis and tools presented here will contribute to understanding interlaminar properties in hierarchical laminate composites with CNT fiber fabrics and unidirectional CF pre-pregs. The intrinsic fracture toughness of CNT veils produced under different processing parameters should clarify the role of veil densification and hopefully lead to methods to modify densified veils, such as those produced semi-industrially, thus paving the way for widespread implementation in structural laminate composites.

**Acknowledgements**

The authors are grateful for generous financial support provided by the European Union Seventh Framework Program under grant agreement 678565 (ERC-STEM), the MINECO (RyC-2014-15115, MAT2015-64167-C2-1-R), the ESTENEA funded by AIRBUS Operations S.L. and CDTI (CIEN 2014 program), and the Cost Action CA15107 (MultiComp). Yunfu Ou appreciates the financial support from the China Scholarship Council [grant number 201606130061].

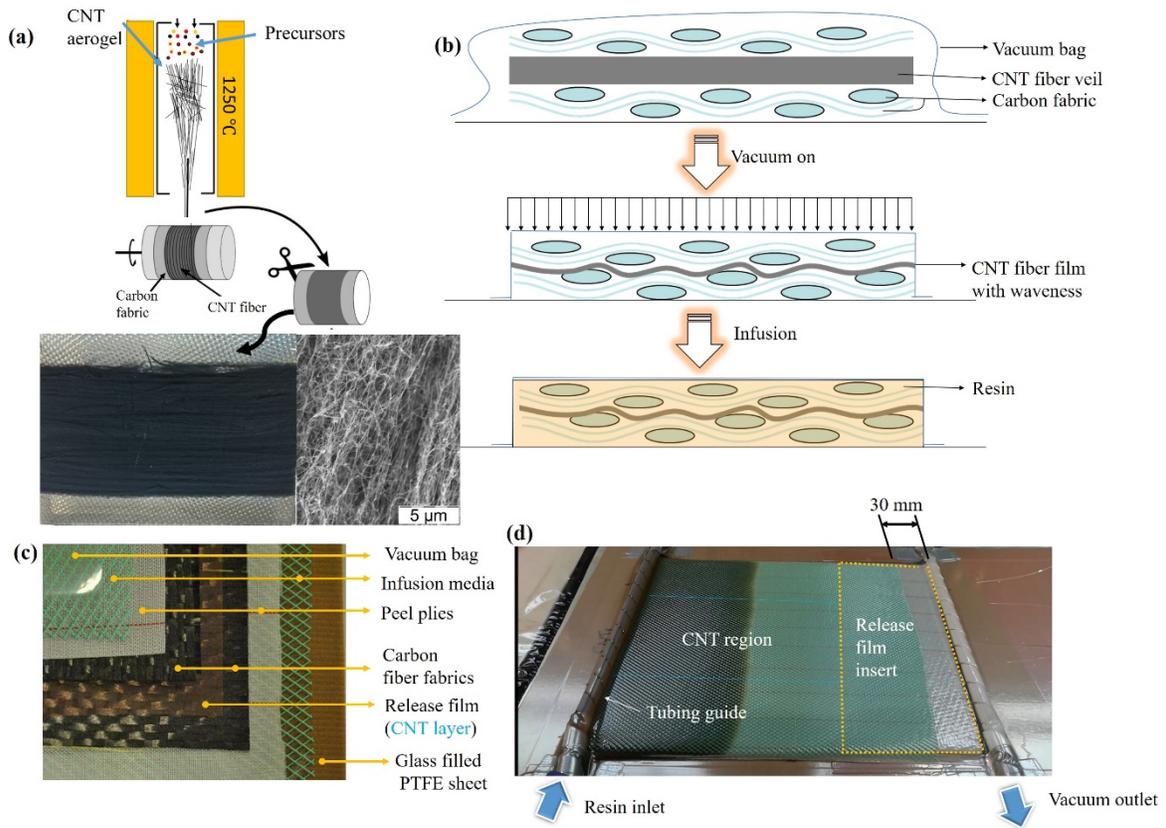

Fig. 1(a) Schematic of direct spinning of CNT fibers from the gas phase by FFCVD; (b) a new protocol for the production of epoxy-based hierarchical composite containing CNT fiber veils used as interleaves; (c) the lay-up of preform and (d) the resin infusion process

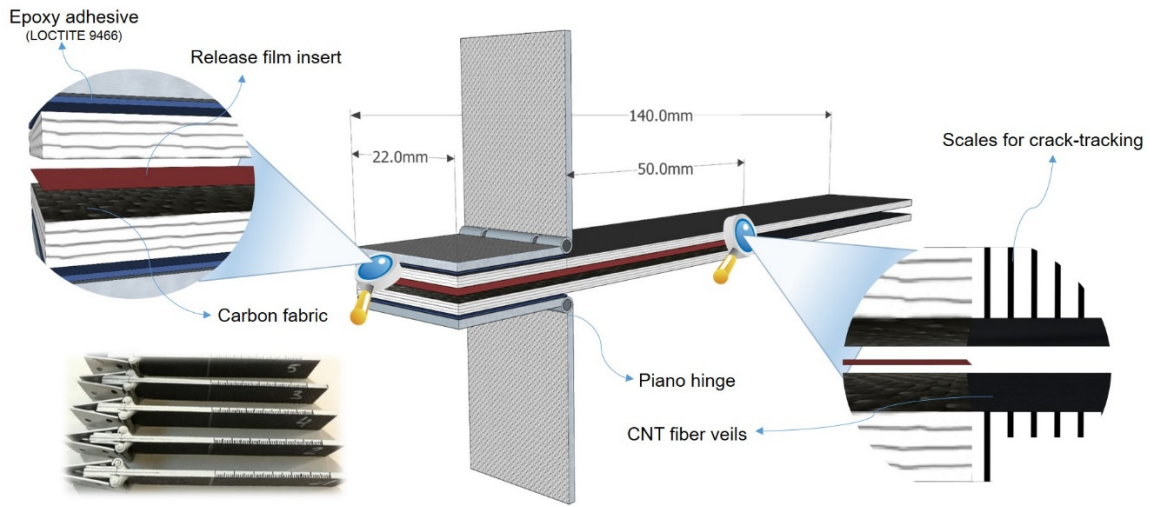

Fig. 2 Schematic configuration of the DCB specimen and a group of real specimens.

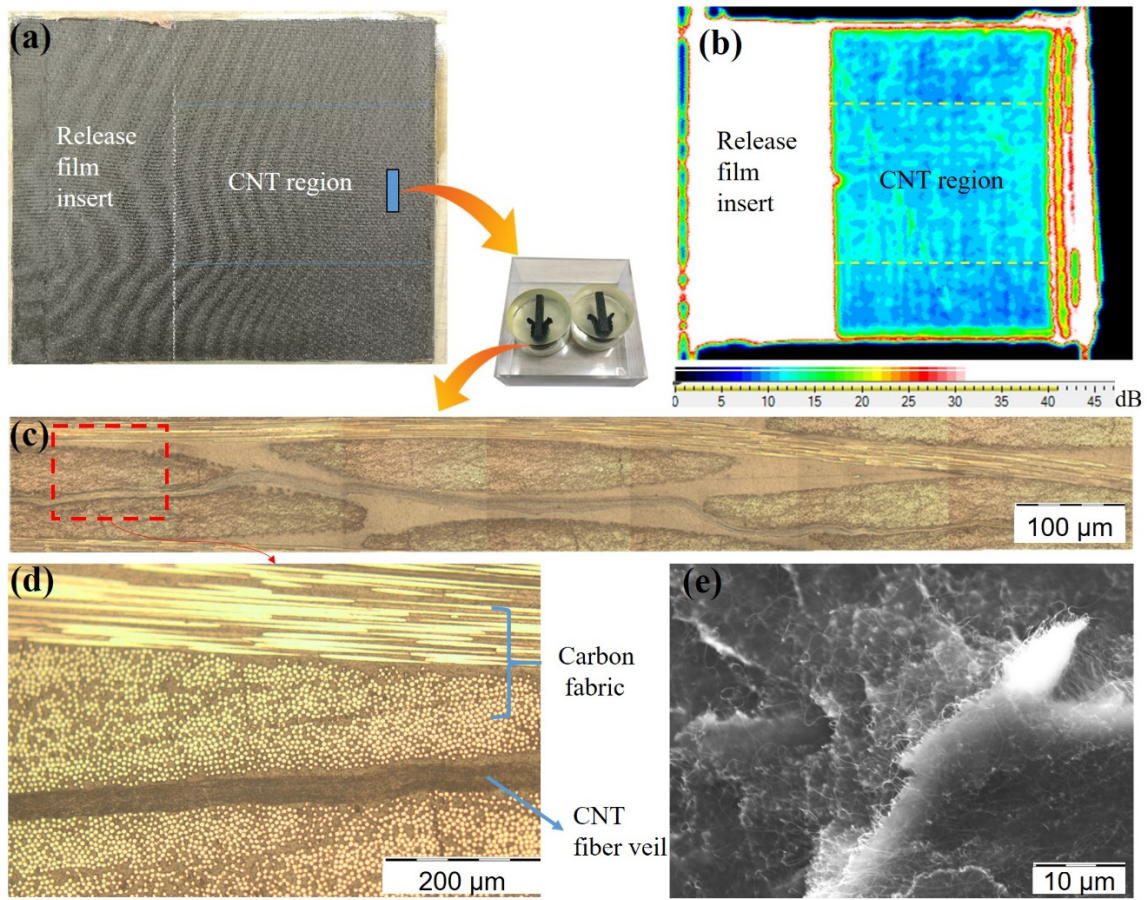

Fig. 3 Characterization of laminates with CNT veils. (a) Surface examination showing optimal surface finish; (b) C-scan picture and (c,d) optical micrographs confirming that there are no appreciable voids in the composite (e) SEM image of CNT veils at the fracture surface.

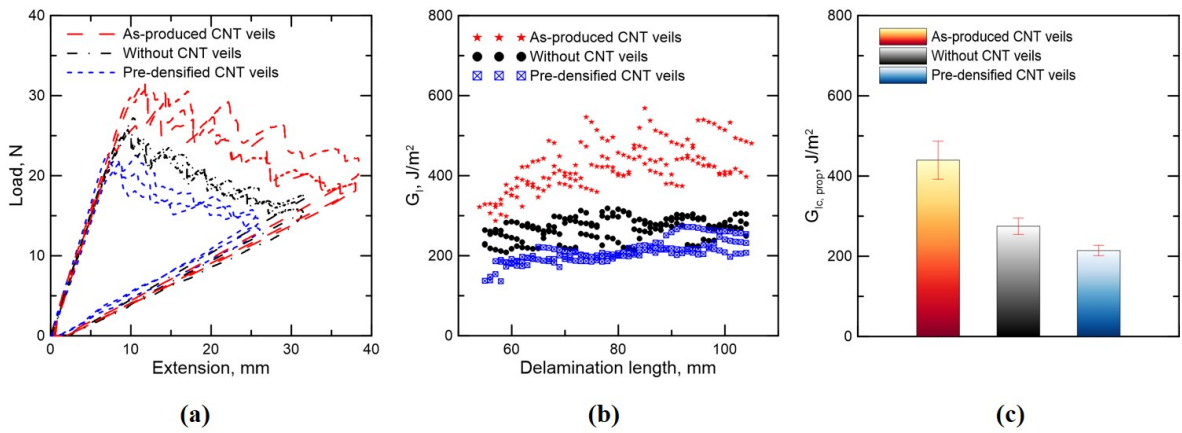

Fig. 4 Representative Load-extension curves (a) and R-curves (b) of specimens with and without CNT veils; (c) comparison of Mode I interlaminar fracture toughness for CNT-interleaved composites (as-produced and pre-densified CNT veils) and baseline (without CNT veils).

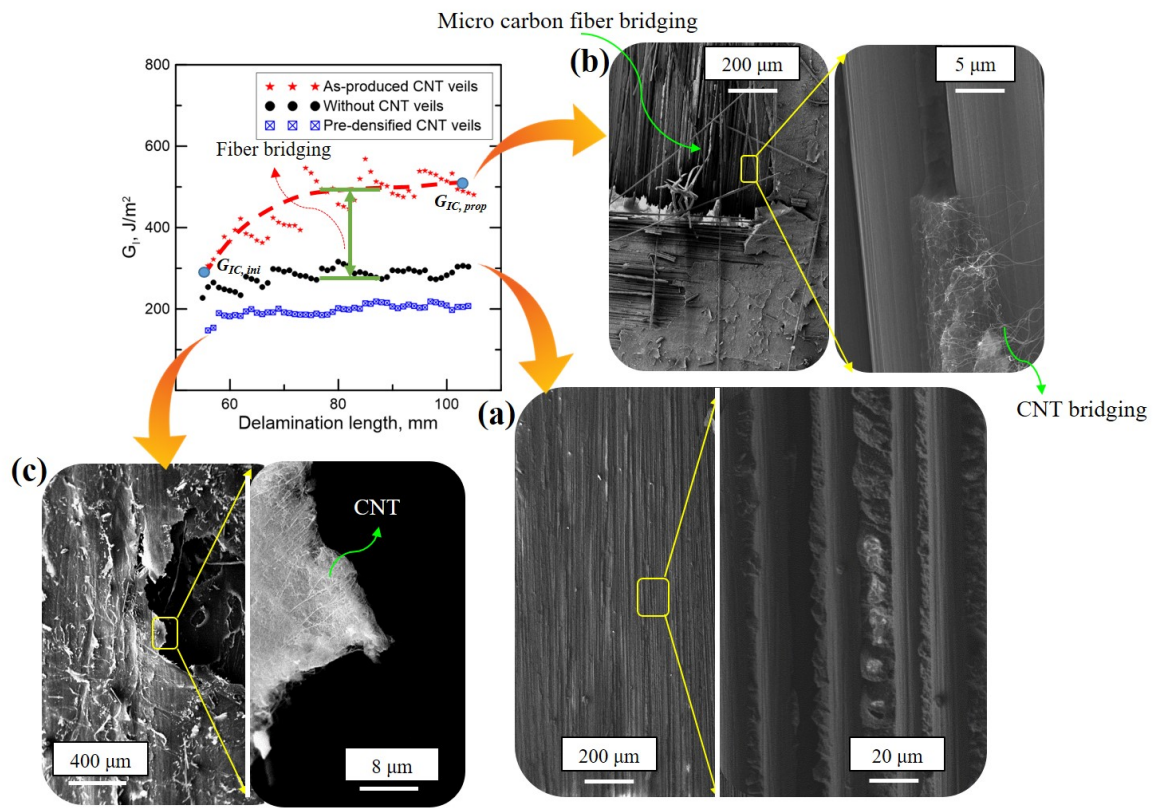

Fig. 5 Comparison of fracture surfaces: (a) Control sample without CNT veils; (b) sample interleaved with as-produced CNT veils and (c) sample containing pre-densified CNT veils.

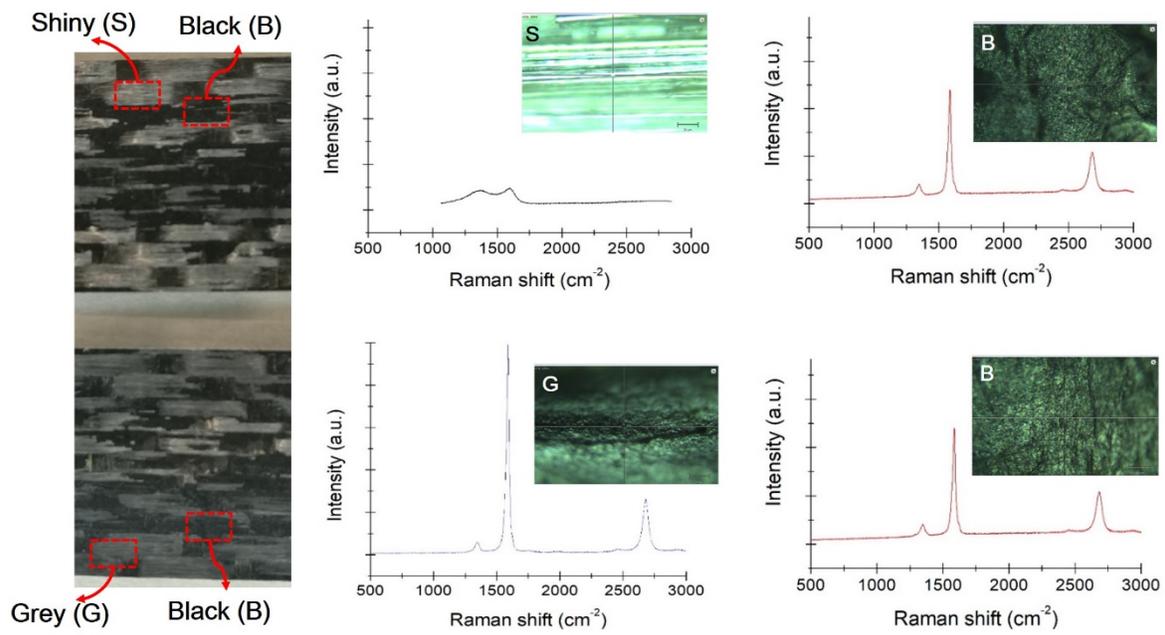

Fig. 6 Three different morphologies (marked as S, B and G) identified on the fracture surface of laminates reinforced with as-produced CNT veils and their Raman spectra.

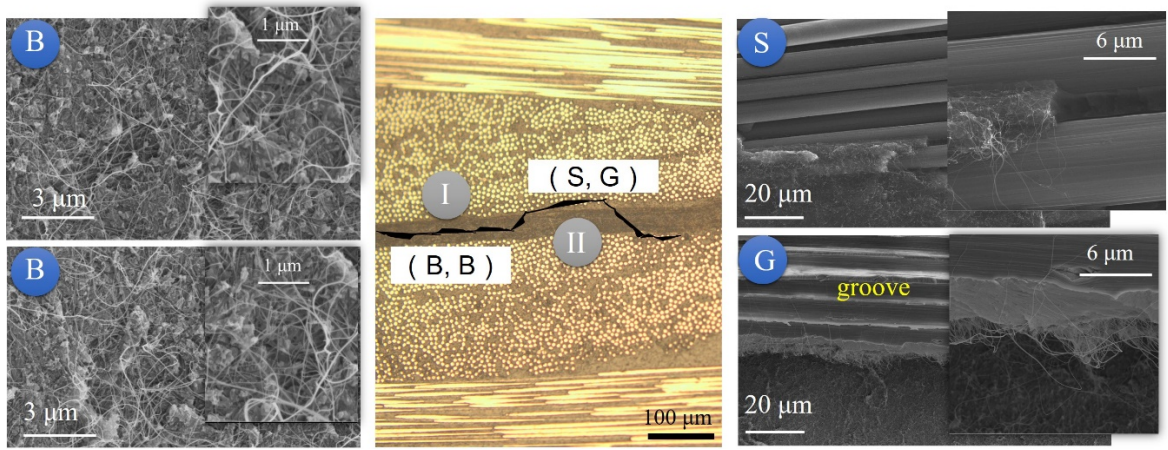

Fig. 7 Crack propagation path determined from combined fractography by Raman spectroscopy, optical and electron microscopy. Visually dark black areas (BB) correspond to cohesive failure of the interleaf. Matching shiny (S) and grey (G) areas represent adhesive failure at the CNT veil/CF interface. Fracture toughness increases when both modes alternate.

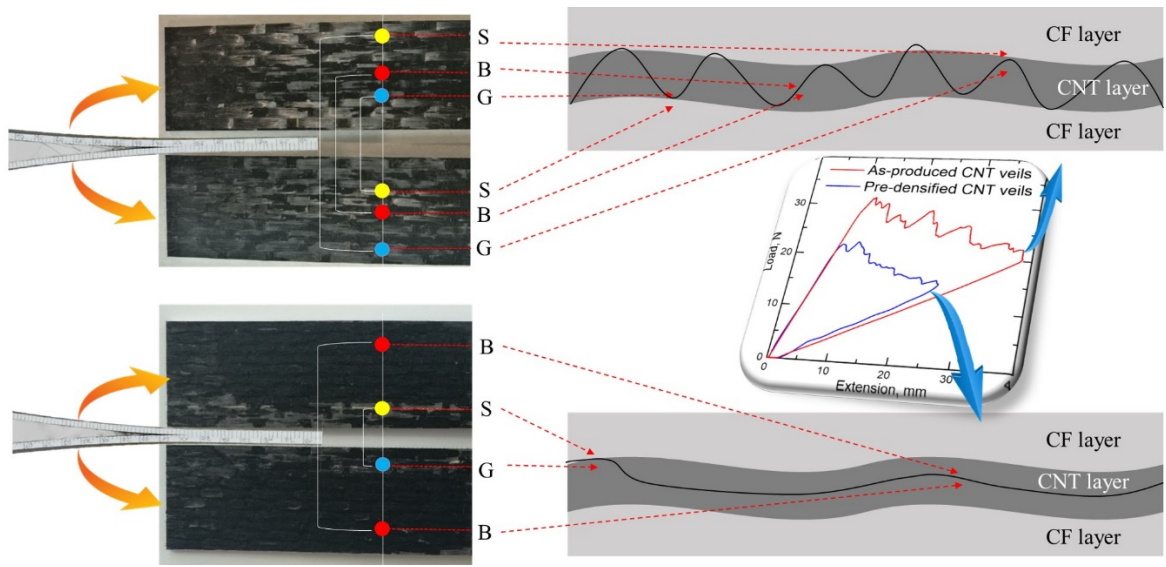

Fig. 8 Fracture surfaces of the two types of composites with CNT veil interleaves. Increases in Mode I fracture toughness correspond to a saw-like crack path alternating between cohesive and adhesive failure in samples with low-density as-produced veils. Pre-densified veils reduced fracture toughness and failed exclusively cohesively.

Tab. 1 Property improvement in hierarchical nanofiller/polymer composites

| Filler integration method | System Nanofiller/fiber/matrix | Fabrication method | Mass fraction/ Area density | CNT geometry | Mechanical improvement | Ref. |
|---|---|---|---|---|---|---|
| Mixing in bulk resin | D&MWCNTs/woven glass/epoxy | RTM | 0.3 wt.% | MWCNT: D =15 nm L=50 μm; DWCNT: D=3 nm L=10 μm | No improvement | [66] |
| | MWCNTs/UD-carbon/epoxy | Autoclave | 1 wt.% | D=10-15 nm L> 500 nm | 60% Mode I 75% Mode II | [67] |
| | S&MWCNTs(-NH$_2$)/UD-carbon/epoxy | Autoclave | 0.5 wt.% | D = 9.5 nm | 75% at initiation and 83% in propagation (Mode I) | [68] |
| | MWCNTs(-COOH)/UD-carbon/epoxy | VARTM | 0.5, 1 and 1.5 wt.% | D=20-30 nm L=10-30 μm | 25%, 20% 17 % at 0.5, 1, 1.5 wt. % (Mode I) | [64] |
| | DWCNTs(-NH$_2$)/woven carbon/epoxy | VARTM | 0.1 wt.% | D=4-7 nm L=1-5 μm | -23% Mode I | [65] |
| In-situ CNT growth | MWCNTs/woven carbon /epoxy | Hand lay-up | - | - | 50% Mode I | [69] |
| | MWCNTs/woven carbon /epoxy | VARTM | 2.24 wt.% | L=10-300 μm | 83% Mode I (Initiation) | [70] |
| | MWCNTs/woven alumina /epoxy | Hand lay-up | - | D=14 nm L=20-30 μm | 76% Mode I | [40] |
| Spray coating | MWCNTs(-COOH)/UD-carbon/epoxy | VARTM | 0.6 wt.% | D=20-40nm<br>L < 5μm | 24% Mode I | [73] |
| | Plasma-modified CNT/UD-glass/epoxy | Hot press | 1.2 g/m² | - | 46% Mode I | [72] |
| | SWCNTs/woven carbon/epoxy | VARTM | 0.01 wt.% | D=1-1.4 nm | 6% Mode I | [75] |
| | CNTs/woven carbon/epoxy | Hand lay-up<br>Vacuum moulding | 0.047 wt.% | - | 50% Mode I | [42] |
| | MWCNTs(-COOH)/UD-carbon/epoxy | Hot press | 0.5 wt.% | D=8-15 nm<br>L=10-50 μm | 17% Mode I | [74] |
| Interleaving | CNT veil/UD-carbon/epoxy | Autoclave | 0.5-2 g/m² | L=1-10 μm | No improvement in Mode I;<br>>50 % Mode II | [48] |
| | CNT veil (-NH₂)/UD-carbon/epoxy | Hand lay-up | 0.2 g/m² | D=10 nm<br>L=300 μm | 13 % Mode I | [76] |
| | Electrospun SPS fibers/UD-glass/epoxy | VARTM | 12-22 g/m² | D=2.1 μm | 90% Mode I<br>100 % Mode II | [24] |
| | Electrospun PA6.9 fibers/woven glass/epoxy | Hot press | 3 g/m²<br>18 g/m² | D=245 nm | 50% for 3 g/m²<br>62% for 18 g/m²<br>(Mode I) | [13] |
| | MWCNT veil/woven carbon/epoxy | VARTM | 0.8 g/m² | D= 3-9 nm | 60% Mode I | This work |

Note: D is diameter of out-layer of CNT or electrospun fiber and d stands for that of inner-layer and L corresponds to the length; AR and CF mean aspect ratio and carbon fiber respectively, and UD- is the abbreviation of unidirectional; –COOH and –NH$_2$ mean that the CNTs were functionalized.